\documentclass{czjphys}         
\begin{document}
\title{Branes, Quantum Nambu Brackets, and the Hydrogen Atom }% use lower case
%
% Leave the remaining items untouched. 
\authori{ Cosmas Zachos}      \addressi{High Energy Physics Division,
Argonne National Laboratory, Argonne, IL 60439-4815, USA \\
zachos@hep.anl.gov}
\authorii{Thomas  Curtright}     \addressii{Department of Physics, 
University of Miami, 
Coral Gables, FL 33124-8046, USA\\
curtright@physics.miami.edu} 
\authoriii{}     \addressiii{} 
%
%Page headings:
\headauthor{ C Zachos and T Curtright}% page heading on the even pages
\headtitle{Quantum Nambu Brackets \ldots} % page heading on the odd pages
\lastevenhead{C Zachos and T Curtright} % p. h. on the last page if even
%%%%%%%%%%%%%% FOR EDITORIAL USE ONLY!!! %%%%%%%%%%%%%%%
\refnum{A}%\total{}\type{} 
\daterec{8/2004} %;\\ final version } 
\issuenumber{11}  \year{2004} 
\setcounter{page}{1}
%\firstpage{1}
%\lastpage{000}
%\makefirsttitle
%%%%%%%%%%%%%%%%%%%%%%%%%%%%%%%%%%%%%%%%%%%%%%%%%%%%%%%%
\maketitle

\begin{abstract} 
The Nambu Bracket quantization of the Hydrogen atom is worked out as an 
illustration of the general method. 
The dynamics of topological open branes is controlled 
classically by Nambu Brackets. Such branes then may be quantized through the
consistent quantization of the underlying Nambu brackets: 
properly defined, the Quantum Nambu Brackets comprise an 
associative structure, although the naive derivation property 
is mooted through operator entwinement. For superintegrable systems, 
such as the Hydrogen atom, the results coincide with those furnished by 
Hamiltonian quantization---but the method is not limited to 
Hamiltonian systems. 
\end{abstract}

This talk by the first author is well-covered by the writeup in 
ref \cite{cincinn}. Here, instead, by way of 
 appendiceal illustration, we briefly extend Pauli's celebrated 
quantization of the Hydrogen atom \cite{pauli} to quantization by  
Quantum Nambu Brackets (QNB) detailed in ref \cite{CQNB}, in 
straightforward application of results in that work.

The classical motion of topological open membranes as well as maximally 
superintegrable systems \cite{cincinn} 
(indeed, most of the maximally symmetric systems solved in 
introductory physics!) 
is controlled by Classical Nambu Brackets (CNB), the multilinear 
generalization of Poisson Brackets (PB) \cite{nambu}. Maximally superintegrable 
systems, are, of course, also described by conventional Hamiltonian mechanics
classically, and are also quantized in standard fashion.

Consider the all-familiar classical Coulomb problem, or, with a view to its 
impending quantization, the Hydrogen-atom problem \cite{pauli}.
Hamilton's equations of motion in phase space are 
\be
\frac{dz^i}{dt} = \{ z^i, H \} , \label{hamilton}
\ee
with $z^i$ standing for the phase-space 6-vector $({\bf r},{\bf p})$, and  
\be
H = \frac{{\bf p}^2}{2}-\frac{1}{r} ~, \label{classicalhamiltonian}
\ee
in simplified (rescaled) notation. 

The invariants of the hamiltonian are the angular momentum vector,
\be
{\bf L} = {\bf r} \times {\bf p},
\ee
and the Hermann-Bernoulli-Laplace vector \cite{goldstein}, 
now usually called the Pauli-Runge-Lenz vector,
\be
{\bf A}=  {\bf p} \times {\bf L} -\hat{{\bf r}}. \label{PRL}
\ee
(Dotting it by $\hat{{\bf r}}$ instantly yields Kepler's elliptical orbits,
$\hat{{\bf r}}\cdot {\bf A}+1 = {\bf L}^2 /r$.)

Since $~{\bf A}\cdot {\bf L}=0$, it follows that 
\be
H=\frac{{\bf A}^2-1}{2{\bf L}^2 } ~~.
\ee
However, to simplify the PB Lie-algebraic structure, 
\be
\{ L_i, L_j \} = \epsilon^{ijk} L_k, \qquad
\{ L_i, A_j \} = \epsilon^{ijk} A_k, \qquad
\{ A_i, A_j \} =-2H  \epsilon^{ijk} L_k, 
\ee
it is useful to redefine 
${\bf D} \equiv \frac{{\bf A}}{\sqrt{- 2 H}}$, and further 
\be  
{\cal R}\equiv {\bf L} + {\bf D}, \qquad {\cal L}\equiv {\bf L} - {\bf D}. 
\ee
These six simplified invariants obey the standard 
$SU(2)\times SU(2) \sim SO(4)$ symmetry algebra, 
\be
\{ {\cal R}_i, {\cal R}_j \} = \epsilon^{ijk} {\cal R}_k, \qquad
\{ {\cal R}_i, {\cal L}_j \} = 0, \qquad
\{ {\cal L}_i, {\cal L}_j \} =\epsilon^{ijk} {\cal L}_k,  \label{su2} 
\ee
and depend on each other and the hamiltonian through  
\be
H=\frac{-1}{2 {\cal R}^2}=\frac{-1}{2 {\cal L}^2} ~, \label{Coulhamiltonian}
\ee
so only five of the invariants are algebraically independent.

Equivalently to the law of motion (\ref{hamilton}), however, the same 
classical evolution
may also be specified by Nambu's equation of motion \cite{nambu}, 
(as is the case for all 
superintegrable systems \cite{nutku}),
\be
\frac{dz^i}{dt} = 
 H^2 ~ \{ z^i, \ln ({\cal R}_3+ {\cal L}_3), 
{\cal R}_1, {\cal R}_2, {\cal L}_1, {\cal L}_2\}~. \label{nambuevolution}
\ee
The object on the right-hand side multiplied by $H^2$ 
is a 6-CNB, i.e. a phase-space Jacobian determinant (volume element),
\be
\{ I_1,I_2,I_3,I_4,I_5,I_6 \}
\equiv {\partial ( I_1,I_2,I_3,I_4,I_5,I_6) 
\over \partial (x,p_x,y,p_y,z,p_z)   } ~.  \label{6CNB}
\ee
It is Nambu's \cite{nambu} celebrated completely antisymmetric multilinear 
generalization of PBs, and, like all even-CNBs, 
it amounts to the Pfaffian \cite{cincinn} of the (antisymmetric) 
matrix with elements $\{ I_i, I_j \}$,
\be
\{ I_1,I_2,I_3,I_4,I_5,I_6 \}= \frac{ \epsilon^{ijklmn} }{48}
\{I_i,I_j\} \{I_k,I_l \} \{I_m,I_n\},   \label{pfaffian}
\ee
i.e., it resolves into a sum of products of PBs, specified uniquely by 
complete antisymmetry and linearity in all arguments $I_i$ \cite{CQNB}.

By utilizing properties of the determinant, such as combining columns,
and Leibniz's rule of differentiation, one easily finds several 
equivalent expressions of 
(\ref{nambuevolution}), as was first worked out in ref \cite{chatterjee}. 
It was suggested in that reference that inclusion of the 
hamiltonian itself among the invariants in the arguments of the CNB might 
be problematic, but, in fact, it is 
quite straightforward: it follows directly from 
(\ref{nambuevolution}) that, alternatively,
\be
\frac{dz^i}{dt} = \frac{1}{4{\cal R}_3 {\cal L}_3 } 
\{ z^i, H ,{\cal R}_1, {\cal R}_2, {\cal L}_1, {\cal L}_2\}~.
\label{tlcalternative} 
\ee
This form and the resolution into PBs (\ref{pfaffian}) serves, by (\ref{su2}), 
 to instantly prove equivalence of (\ref{nambuevolution}) to (\ref{hamilton}). 
Motion is evidently confined on the constant surfaces specified by 
the five invariants entering in the CNB---a generic 
feature of the CNB description of maximally superintegrable systems 
\cite{cincinn,CQNB,nutku,chatterjee,tegmen}; any algebraically independent 
invariants would do.
 
Actually, this problem has already been addressed in the treatment of $S^3$ in
ref \cite{CQNB}, eqns (56,61), except that the respective hamiltonian 
in that problem is the inverse of the present Coulomb one.

The action whose extremization yields this evolution law is a topological
5-form action 
\be
S=\int \Bigl ( x~ dp_x \wedge dy \wedge dp_y \wedge dz \wedge dp_z
+ \ln ({\cal R}_3+ {\cal L}_3) ~d{\cal R}_1\wedge d{\cal R}_2
\wedge  d{\cal L}_1 \wedge  d{\cal L}_2 \wedge dt\Bigr ) ~,
\ee
a Cartan integral invariant ``4-brane" action analogous to $(4+1)$-dimensional 
$\sigma$-model  WZWN topological interaction terms 
\cite{cincinn,estabrook}.  
It originates by Stokes' law in the integral on an open 6-surface of the 
exact  6-form 
\begin{eqnarray}
d\omega_5 = dx\wedge  
dp_x \wedge dy \wedge dp_y \wedge dz \wedge dp_z+
d\ln ({\cal R}_3+ {\cal L}_3) \wedge d{\cal R}_1\wedge d{\cal R}_2
\wedge  d{\cal L}_1 \wedge  d{\cal L}_2 \wedge dt &&\\
=\left (dx-\{ x , \ln ({\cal R}_3+ {\cal L}_3), {\cal R}_1, {\cal R}_2,
{\cal L}_1 ,{\cal L}_2 \} dt\right )\wedge \left 
(dp_x-\{ p_x , \ln ({\cal R}_3+ {\cal L}_3), {\cal R}_1, {\cal R}_2,
{\cal L}_1, {\cal L}_2 \} dt\right )&& \nonumber \\ 
\wedge   \left (dy-\{ y, \ln ({\cal R}_3+ {\cal L}_3), {\cal R}_1, {\cal R}_2,
{\cal L}_1 ,{\cal L}_2 \} dt\right )\wedge \left 
(dp_y-\{ p_y , \ln ({\cal R}_3+ {\cal L}_3), {\cal R}_1, {\cal R}_2,
{\cal L}_1, {\cal L}_2 \} dt\right )&& \nonumber \\
\wedge\left (dz-\{ z , \ln ({\cal R}_3+ {\cal L}_3), {\cal R}_1, {\cal R}_2,
{\cal L}_1 ,{\cal L}_2 \} dt\right )\wedge \left 
(dp_z-\{ p_z , \ln ({\cal R}_3+ {\cal L}_3), {\cal R}_1, {\cal R}_2,
{\cal L}_1, {\cal L}_2 \} dt\right ) .&& \nonumber    
\end{eqnarray}

For any function of phase space with no explicit time dependence, then, 
the classical evolution law 
\be
\frac{df}{dt} = 
 H^2 \{ f, \ln ({\cal R}_3+ {\cal L}_3), 
{\cal R}_1, {\cal R}_2, {\cal L}_1, {\cal L}_2\}  \label{classevol}
\ee
is to be quantized 
($\hbar$-deformed) consistently below, as detailed for $S^3$ 
in refs \cite{cincinn,CQNB}.

As noted by Pauli, extension to operators requires a hermitean version of  
(\ref{PRL}), 
\be
{\bf A}'=\sfrac{1}{2} ({\bf p}\times{\bf L}-{\bf L}\times{\bf p})-\hat{{\bf r}}, 
\ee
so that 
\be
({\bf A}')^2=2 H ( {\bf L}^2 + \hbar^2) + 1 ,
\ee
leading to ${\bf D}' \equiv \frac{{\bf A}'}{\sqrt{- 2 H}}$, and further 
to the respective chiral reduction ${\cal R}'$ and ${\cal L}'$, which 
obey 
\be
[ {\cal R}'_i, {\cal R}'_j ]=2i\hbar ~ \epsilon^{ijk} {\cal R}'_k, \qquad
[ {\cal R}'_i, {\cal L}'_j ]=0 ,\qquad
[ {\cal L}'_i, {\cal L}'_j ]=2i\hbar ~ \epsilon^{ijk} {\cal L}'_k,
\label{qsu2} \ee
and hence 
\be
H=\frac{-1}{2 ({\cal R}'^2 +\hbar^2)}=
\frac{-1}{2 ({\cal L}'^2 +\hbar^2)}~.
\ee
One may thus omit the primes on the operator expressions without 
appreciable loss of clarity, and recall the eigenvalues of the 
quadratic Casimir invariants of $SU(2)$ for $s=0, \frac{1}{2}, 1,...$,
leading to the Balmer spectrum for the hamiltonian, 
\be
\langle H \rangle = \frac{-1}{2\hbar^2 (4s(s+1)+1)} =
\frac{-1}{2\hbar^2 (2s+1)^2}                        ~.
\ee
The size of these $SU(2)\times SU(2)$ multiplets, $(2s+1)^2$, is the 
corresponding degeneracy.
 
Time evolution in the hamiltonian picture is given by Heisenberg's quantum 
equation of motion, 
\be 
i\hbar \frac {df}{dt} = [f,H] ~;
\ee
in the QNB picture (detailed in refs \cite{cincinn,CQNB}) it works as 
outlined below.

A 6-QNB, $[I_1,I_2,I_3,I_4,I_5,I_6 ]$, consists of of the fully antisymmetrized
linear product of its 6 operator arguments. Analogously to its 6-CNB 
counterpart,
it can be shown to resolve to a sum of strings of commutators, 
\be
[ I_1,I_2,I_3,I_4,I_5,I_6 ] = \frac{ \epsilon^{ijklmn} }{8}
[I_i,I_j ]   [ I_k,I_l]  [I_m,I_n] . \label{Qpfaffian}
\ee
This is longer than its classical counterpart
(which has only $15=5\cdot 3$ distinct terms), as commutators need not
commute with each other in general, so their symmetric entwinement leads to 
$90=3! 5\cdot 3$ terms. In practical terms, in general, evaluation of even 
QNBs resolves to judicious evaluation of commutators \cite{CQNB}. 
It is evident by inspection of this resolution that the classical limit 
($\hbar \rightarrow 0$) of this QNB is the CNB discussed,
\be
[ I_1,I_2,I_3,I_4,I_5,I_6 ] ~ \rightarrow ~ 3!(i\hbar)^3 ~ 
\{ I_1,I_2,I_3,I_4,I_5,I_6\} .
\ee 

It follows then from (\ref{Qpfaffian}) (cf eqns (183)+(184) of \cite{CQNB}) 
that, for the particular 
simple Lie algebras (\ref{qsu2}), the quantization of (\ref{classevol})
is just
\be
\! 3(i\hbar)^3\!\left ( ({\cal R}_3 +{\cal L}_3) \frac{df}{dt} + 
\frac{df}{dt} ({\cal R}_3 + {\cal L}_3) \right ) \! = \!
H \Bigl [ f, {\cal R}_3+ {\cal L}_3,{\cal R}_1, {\cal R}_2, {\cal L}_1, 
{\cal L}_2 \Bigr ]  H + {\cal Q}(O(\hbar^5)).
\ee
${\cal Q}(O(\hbar^5))$ is a subdominant nested commutator 
``quantum rotation" \cite{CQNB}, vanishing in the classical limit\footnote{
Specifically, ${\cal Q}= 2\hbar^2 H \sum_i 
([[[f, {\cal L}_i],{\cal L}_i], {\cal R}_3]+
[[[f, {\cal R}_i],{\cal R}_i], {\cal L}_3])  H$.}.
Solving for $df/dt$ may be more challenging technically 
(the Jordan-Kurosh spectral problem), 
but the formulation is still equivalent to the standard Hamiltonian 
quantization of this problem \cite{CQNB}. Of course, expectation values 
in sectors with definite $L_3=({\cal L}_3 +{\cal R}_3)/2$ are thus
proportional to $\langle \frac{df}{dt}  \rangle$.

Similarly, (cf eqn (77) of ref \cite{CQNB}, also for $S^3$),
equivalent forms such as (\ref{tlcalternative}) quantize through
\be
4(i\hbar)^3 \left   ( {\cal R}_3, {\cal L}_3, \frac{df}{dt} \right ) =
 \Bigl [ f, H,{\cal R}_1, {\cal R}_2, {\cal L}_1, {\cal L}_2 \Bigr ],
\ee
where the parenthesis on the left indicates the complete symmetrization of
its three arguments.

In contrast to Heisenberg's law of motion in the Hamiltonian formulation,
the operator acting on $f$ in the QNB formulation above is not a 
derivative operator, i.e., it does not obey Leibniz's chain rule, because 
the actual (time) derivatives are entwined with other operators. This was 
all too widely thought 
to be an obstacle in utilizing QNBs which are not derivative operators 
themselves, but it was shown to not be a consistency  
problem at all, for even-QNBs\footnote{Odd-QNBs may be defined consistently 
through even ones of higher rank.} \cite{CQNB,cincinn}.
In quantization, associativity trumps naive derivation features\footnote
{As a formal wisecrack, consider 
defining the following antisymmetric bracket for a fixed operator $0$,
$[\! [ A,B ]\! ]\equiv AOB-BOA$. This clearly satisfies the Jacobi identity, 
but it fails a naive derivation property, as 
$[\![ A,BC ]\! ]\neq [\![ A,B ]\! ] C + B [\![ A,C ]\! ]$. 
Of course, suitable insertions of $O$ in all products would restore that
property here.}.

This general methodology has proven successful in a large
number of systems, including some non-Hamiltonian ones \cite{cincinn}.

\bigskip  
\noindent{\small Thanks are due to C Burdik and the 
organizers of the International Colloquium. This work was supported by the US 
Department of Energy, Division of High Energy Physics, 
Contract W-31-109-ENG-38, and the NSF Award 0303550.}

\end{document}